\documentclass[10pt,twocolumn,letterpaper]{article}

\usepackage{iccv}
\usepackage{times}
\usepackage{epsfig}
\usepackage{graphicx}
\usepackage{amsmath}
\usepackage{amssymb}
\usepackage{romannum}

\usepackage[breaklinks=true,bookmarks=false]{hyperref}

\iccvfinalcopy 

\def\iccvPaperID{****} 

\ificcvfinal\fi

\begin{document}

\title{W-Net: Two-stage U-Net with misaligned data for raw-to-RGB mapping}

\author{Kwang-Hyun Uhm, Seung-Wook Kim, Seo-Won Ji, Sung-Jin Cho, Jun-Pyo Hong, Sung-Jea Ko\thanks{Corresponding author}\\
School of Electrical Engineering, Korea University\\
Seoul, Korea\\
{\tt\small \{khuhm, swkim, swji, sjcho, jphong\}@dali.korea.ac.kr, sjko@korea.ac.kr}
}

\maketitle
\ificcvfinal\thispagestyle{empty}\fi

\begin{abstract}
   Recent research on learning a mapping between raw Bayer images and RGB images has progressed with the development of deep convolutional neural network. A challenging data set namely the Zurich Raw-to-RGB data set~(ZRR) has been released in the AIM 2019 raw-to-RGB mapping challenge. In ZRR, input raw and target RGB images are captured by two different cameras and thus not perfectly aligned. Moreover, camera metadata such as white balance gains and color correction matrix are not provided, which makes the challenge more difficult. In this paper, we explore an effective network structure and a loss function to address these issues. We exploit a two-stage U-Net architecture, and also introduce a loss function that is less variant to alignment and more sensitive to color differences. In addition, we show an ensemble of networks trained with different loss functions can bring a significant performance gain. We demonstrate the superiority of our method by achieving the highest score in terms of both the peak signal-to-noise ratio and the structural similarity and obtaining the second-best mean-opinion-score in the challenge.
\end{abstract}

\section{Introduction}

In this paper, we describe our solution for the AIM 2019 challenge on raw-to-RGB mapping \cite{AIM2019RAW}. The challenge releases a Zurich Raw-to-RGB~(ZRR) data set for the task. The ZRR data set consists of pairs of raw and RGB images which are captured by Huawei P20 and Canon 5D Mark \Romannum{4} cameras, respectively. The challenge aims at learning mapping between input raw and target RGB images based on the image pairs given in ZRR data set. However, in this data set, the input and target images are not perfectly aligned as they are taken with different cameras. Moreover, some camera metadata such as white balance gains and color correction matrix are not provided, which makes the task more difficult.

In general, digital cameras process the raw sensor data through the image processing pipeline to produce the desired RGB images. A traditional camera imaging pipeline includes a sequence of operations such as white balance, demosaicing, denoising, color correction, gamma correction, and tone mapping. Typically, each operation is performed independently and requires hand crafted priors. With the recent advances in deep convolutional neural networks~(CNNs), research on implementing the imaging pipeline using CNN also has progressed. Schwartz~\textit{et~al.}~\cite{schwartz2018deepisp} proposed a CNN architecture named DeepISP to perform end-to-end image processing pipeline. DeepISP achieved better visual quality scores than the manufacturer image signal processor. Chen~\textit{et~al.}~\cite{chen2018learning} developed a CNN to learn the imaging pipeline for short-exposure low-light raw images. However, these methods trained the network using aligned pairs of input raw and target RGB images obtained by the same camera. 

Some studies have attempted to convert the image captured by one camera to the image taken by another camera. Nguyen~\cite{nguyen2014raw} proposed a calibration method to find a mapping between two raw images from the two different cameras. Ignatov~\cite{ignatov2017dslr} proposed a CNN-based method of learning a mapping from mobile camera images into DSLR images using RGB image pairs. However, these methods only handle the mapping between the images in the same color space.

In this work, we explore an effective network structure and loss functions to address the challenging issues in ZRR. We exploit a two-stage U-net architecture with network enhancements. As U-net utilizes features that are down-sampled several times, these features are relatively invariant to small translation and rotation of the contents in an image. To extract more informative features for our task, we employ a channel attention mechanism. Specifically, we only apply the channel attention module to the expanding path of U-Net. In the expanding path, features are up-sampled and combined with the high-resolution features from the contracting path. Then, the combined features are channel-wise weighted according to the global statistics of the activation to contain more useful information. We also add a long skip connection to U-Net to ease the training of the network. Our experiment demonstrates that the performance on ZRR can be improved by these network enhancements. Though single enhanced U-Net achieves comparable performance, we exploit two-stage U-Net architecture to further boost the performance in the challenge. We cascade the same enhanced U-Net to refine the output RGB images of the first stage.

Also, we introduce a loss function that is less variant to the alignment of training data and encourages the network to generate well color-corrected images. We utilize the perceptual loss \cite{johnson2016perceptual} to handle the misalignment between input raw and target RGB images. We use high-level features from a deep network since they are down-sampled multiple times and thus effective for learning with misaligned data. Since the color correction step is implemented in a camera image processing pipeline, the network needs to inherently learn this step to reconstruct RGB images. To encourage the network to learn an accurate color transformation, we introduce a color loss which is defined by the cosine distance between the RGB vectors of predicted and target images. 

Finally, we apply the model ensemble method to improve the quality of output images. Unlike the typical model ensemble method, we trained the networks with different loss functions and averaged the outputs. Our experiment shows that the ensemble of three networks trained with three different loss functions brings significant improvement of performance. 
We achieved the best performance in terms of the peak signal-to-noise ratio~(PSNR) of 22.59 dB and the structural similarity~(SSIM) of 0.83 in the AIM 2019 raw-to-RGB mapping challenge - Track 1: fidelity, and the second-best performance in Track 2: perceptual.

\section{Related work}

\paragraph{Image signal processing pipeline.}
There exist various image processing sub-tasks inside the traditional ISP pipeline. The most representative method includes denoising, demosaicing, white balancing, color correction, and tone mapping~\cite{brooks2019unprocessing,ramanath2005color}. The demosaicing operation interpolates the single-channel raw image with repeated mosaic patterns into multi-channel color images~\cite{gharbi2016deep}. Denoising operation removes the noise occurred in a sensor and enhances the signal-to-noise ratio~\cite{dabov2007image}. White balancing step corrects the color shifted by illumination according to human perception\cite{Cheng_2015_ICCV}. Color correction applies a matrix to convert color space of the image from raw to RGB for display~\cite{brooks2019unprocessing}.
Tone mapping compresses the dynamic range of the raw image and enhances the image details~\cite{yuan2012automatic}

In the traditional image processing pipeline, each step is designed using the handcrafted priors and performed independently. This may cause an error accumulation when processing the raw data through the pipeline~\cite{heide2014flexisp}. 

\begin{figure}[t]
\centering 
\includegraphics[width=1.0\linewidth]{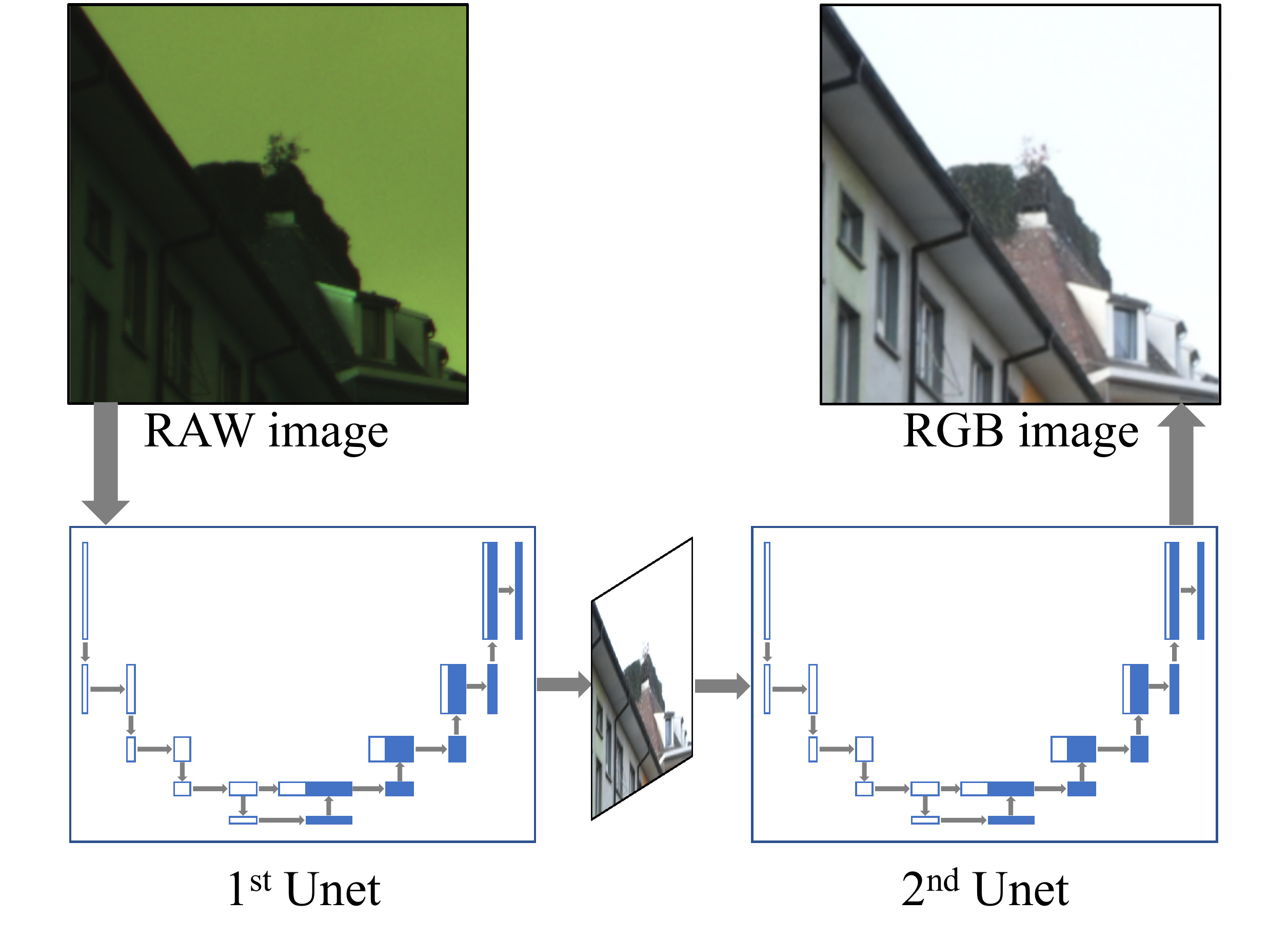}
\DeclareGraphicsExtensions.
\caption{Illustration of the process of the W-Net.}
\label{fig:1}
\end{figure}

\paragraph{Deep learning on imaging pipeline.}
As CNNs have shown significant success in low-level image processing tasks, such as demosaicing~\cite{gharbi2016deep}, denoising~\cite{brooks2019unprocessing,guo2019toward,zhang2017beyond}, deblurring~\cite{sun2015learning}, some studies~\cite{buckler2017reconfiguring,chen2018learning,schwartz2018deepisp} utilize CNNs to model the camera imaging pipeline. Schwartz~\etal\ proposed a CNN model to perform demosaicing, denoising and image enhancement together~\cite{schwartz2018deepisp}. 
Chen ~\etal\ developed a CNN to learn the imaging pipeline for low-light raw images~\cite{chen2018learning, shen2017msr}. 

Converting image taken by one camera to another camera has also been studied.
Nguyen~\etal\ proposed a calibration method to obtain raw-to-raw mapping between image sensor color responses~\cite{nguyen2014raw}.
Ignatov~\etal\ proposed a method to learn the mapping between images taken by mobile phone and a DSLR camera. However, they use, as an input, an image already processed by an image signal processor~\cite{ignatov2017dslr}.

\begin{figure*}[t]
\centering 
\includegraphics[width=1.0\linewidth]{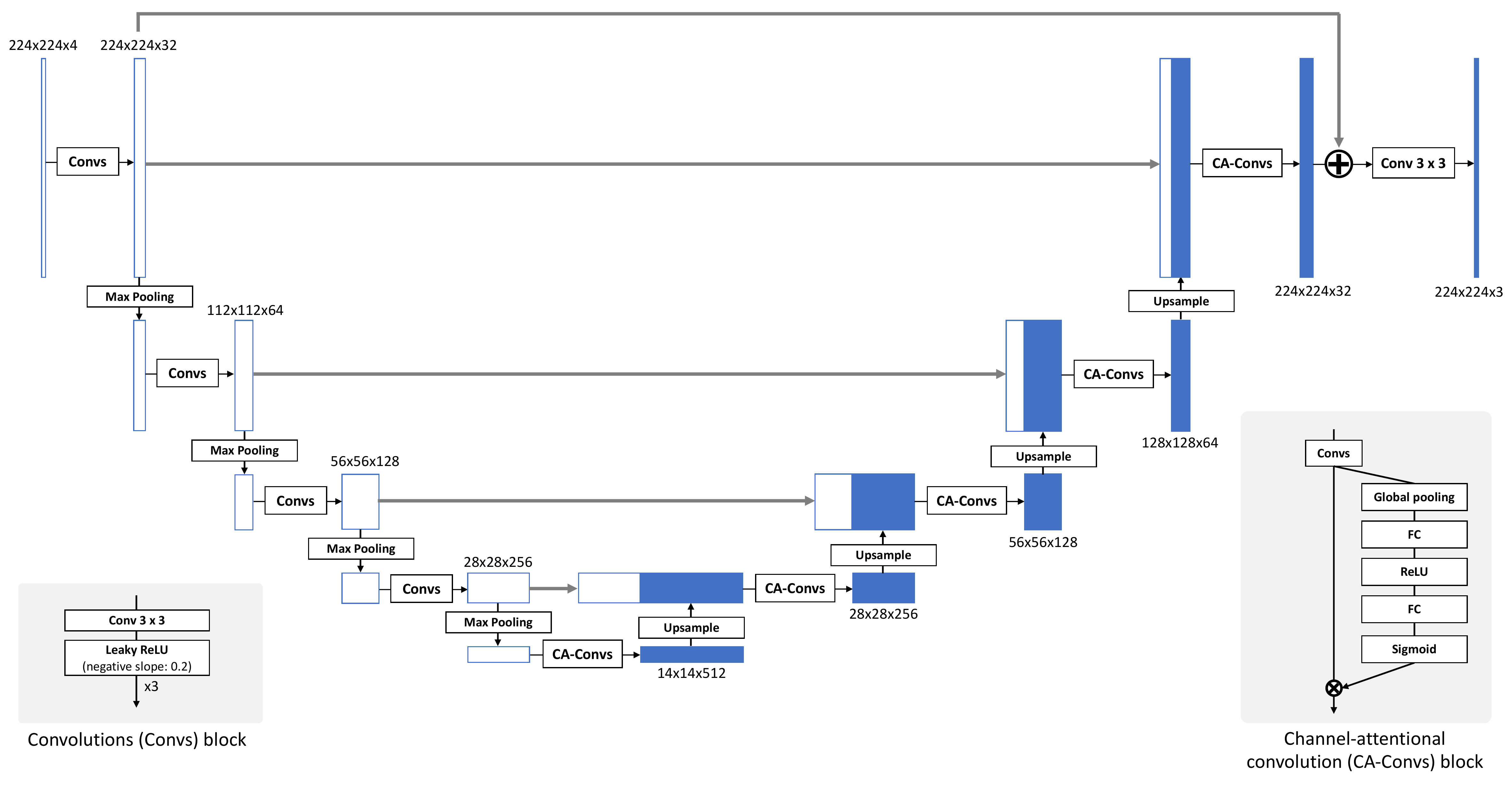}
\DeclareGraphicsExtensions.
\caption{The architecture of the enhanced U-Net. Best viewed with zoom.}
\label{fig:2}
\end{figure*}

\section{Methodology}
\subsection{Network Architecture}

Figure~\ref{fig:1} shows our two-stage U-Net based network, called W-Net, for raw-to-RGB mapping. We utilize the U-Net \cite{Olaf2015unet} because the structure consisting of multiple pooling and un-pooling layers is effective for learning on the misaligned data. In W-Net, the RGB image is first reconstructed by the single U-Net based network and then refined by the cascaded network.   

We enhance the U-Net for our task. Figure~\ref{fig:2} shows the architecture of the enhanced U-Net.
U-Net consists of a contracting path and an expanding path.
In the contracting path, we apply convolutions (Convs) blocks and 2$\times$2 max-pooling operations to extract and down-sample features, where the Convs block consists of three 3$\times$3 convolution layers followed by parametric rectified linear units (PReLU) \cite{He_2015_ICCV} with negative slope of 0.2.
Note that the number of channels of the features is doubled at each Convs block. 
In the expanding path, features are up-sampled by bilinear interpolation and concatenated with the features from the contracting path of the same size. To obtain more informative features for our task, we employ channel attention mechanism. Specifically, the channel attentional-convolutions block (CA-Convs) block is applied at each step of the expanding path. In the CA-Convs block, output features of the Convs block are first global average pooled to take the global spatial information and then transformed by fully connected layers (FC) and ReLU to represent the channel importance. The weights are obtained by the sigmoid function and multiplied to the outputs of the Convs block. The use of CA-Convs block largely improves the performance with a negligible parameter increase (see Sec.~\ref{abstudy}).  Note that applying the CA-Convs blocks to both the contracting and expanding path does not increase the performance in our experiment. We also add a long skip connection between the highest resolution features to ease the training of the network. After the long skip connection, a $3 \times3$ convolution is performed to produce the RGB image.

\subsection{Loss Function}
In this section, we describe our loss function which consists of three terms. we denote $T$ as the target RGB image and $\hat{T}$ as the predicted image.

\paragraph{Pixel loss.}
First, we adapt the pixel-wise $L_{1}$ loss, which is defined by ${L_{pixel} = ||\hat{T} - T||_1}$. However, using only the pixel loss leads to blurry results because image pairs are misaligned (see Sec.~\ref{abstudy}). 

\paragraph{Feature loss.}
To handle data misalignment, we utilize the perceptual loss function \cite{johnson2016perceptual}. As features that are down-sampled multiple times by pooling layers are less sensitive to the mild misalignment of images, we extract the high-level features from the pretrained VGG-19 network \cite{simonyan2014very}  and calculate $L_{1}$ distance between the extracted features of $\hat{T}$ and $T$. Therefore, our feature loss can be written as:  
\begin{equation}
L_{feat} = || \phi (\hat{T}) - \phi (T) ||_1 ,
\end{equation}
where $\phi$ denotes the `relu4\_1' or `relu5\_1' feature of the VGG-19 network. By using the feature loss, we could obtain less blurry output images with fine details (see Figure~\ref{fig:3}).

\paragraph{Color loss.}
We further define the color loss to learn an accurate color transformation between input raw and target RGB images. We measure the cosine distance between the RGB color vectors of the down-sampled predicted and target image. The color loss can be written as:

\begin{equation}
L_{color} = 1- \frac{1}{N} \sum_{i}^{N} \frac {\pmb {\hat{T}}^{\downarrow_{2}}_{i} \cdot \pmb {T}^{\downarrow_{2}}_{i}}{||\pmb {\hat{T}}^{\downarrow_{2}}_{i}|| \   ||\pmb {T}^{\downarrow_{2}}_{i}||},
\end{equation}
where $\cdot$ is the inner product operator, $\downarrow_2$ denotes down-sampling operator by a factor of 2, and $\hat{T}^{\downarrow_{2}}_{i}$ and ${T}^{\downarrow_{2}}_{i}$ are the $i^{th}$ RGB pixel values of  $\hat{T}^{\downarrow_{2}}$ and $T^{\downarrow_{2}}$, respectively.

Finally, we define our loss function by the sum of the aforementioned losses as follows:
\begin{equation}
L_{total} = L_{pixel} + L_{feat} + L_{color}.
\end{equation}

\section{Experiments}

\subsection{Dataset and Training Details}
\paragraph{Dataset.}
ZRR provides 89,000 pairs of raw and corresponding RGB image with the size 224$\times$224, where raw and RGB images are taken by Huawei P20 and Canon 5D Mark \Romannum{4}, respectively. We used 88,000 image pairs for training our model and the remaining 1,000 pairs for validation. We normalized and denormalized the input and the predicted images, respectively, by the mean and standard deviation of the whole training data. Data augmentations such as flipping and rotation were not applied.

\paragraph{Training details.}
We implemented our model using Pytorch framework with Intel i7, 32GB of RAM, and NVIDIA Titan XP. Mini-batch size was set to 24. We trained our model using Adam optimizer \cite{kingma2014adam} with $\beta_1$=0.9, $\beta_2$=0.999.  The first U-Net of our model was trained for 100 epochs and then the weights of the first network were frozen. Then, the second U-Net was trained for 25 epochs. Learning rate was initialized to 10$^{-4}$ and dropped to 10$^{-5}$ at the last one epoch. Approximately 3 days were required to train our model.

\subsection{Ablation Study}
\label{abstudy}
\paragraph{Network architecture.}
First, we demonstrated the effectiveness of our network model. The experimental results are shown in Table 1. We trained models using only the pixel loss described in Sec. 3.2. Our basic network is a single original U-Net. By applying channel attention (CA) modules, an improvement of 0.2 dB was obtained. Also, adding the long skip connection (LSC) increased the PSNR by 0.2 dB. Note that these improvements were achieved with negligible parameter increase. In addition, cascading the same enhanced U-Net brought 0.1 dB performance gain. These results suggest that our network design is effective for learning the raw-to-RGB mapping task.

\begin{table}[t]
\renewcommand{\arraystretch}{1.2}
\label{table:1}
\begin{center}
\begin{tabular}{c||c|c|c|c}
\hline \hline
CA &  & \checkmark & \checkmark & \checkmark \\
LSC &  &  & \checkmark & \checkmark  \\
Two-stage &  &  &  & \checkmark \\
\hline
PSNR & 22.30 & 22.58 & 22.72 & 22.82 \\
SSIM & 0.8687 & 0.8704 & 0.8700 & 0.8741 \\
\hline \hline
\end{tabular}
\end{center}
\caption{Ablation studies on network architectures}
\end{table}

\begin{figure*}[t]
\centering 
\includegraphics[width=1.0\linewidth]{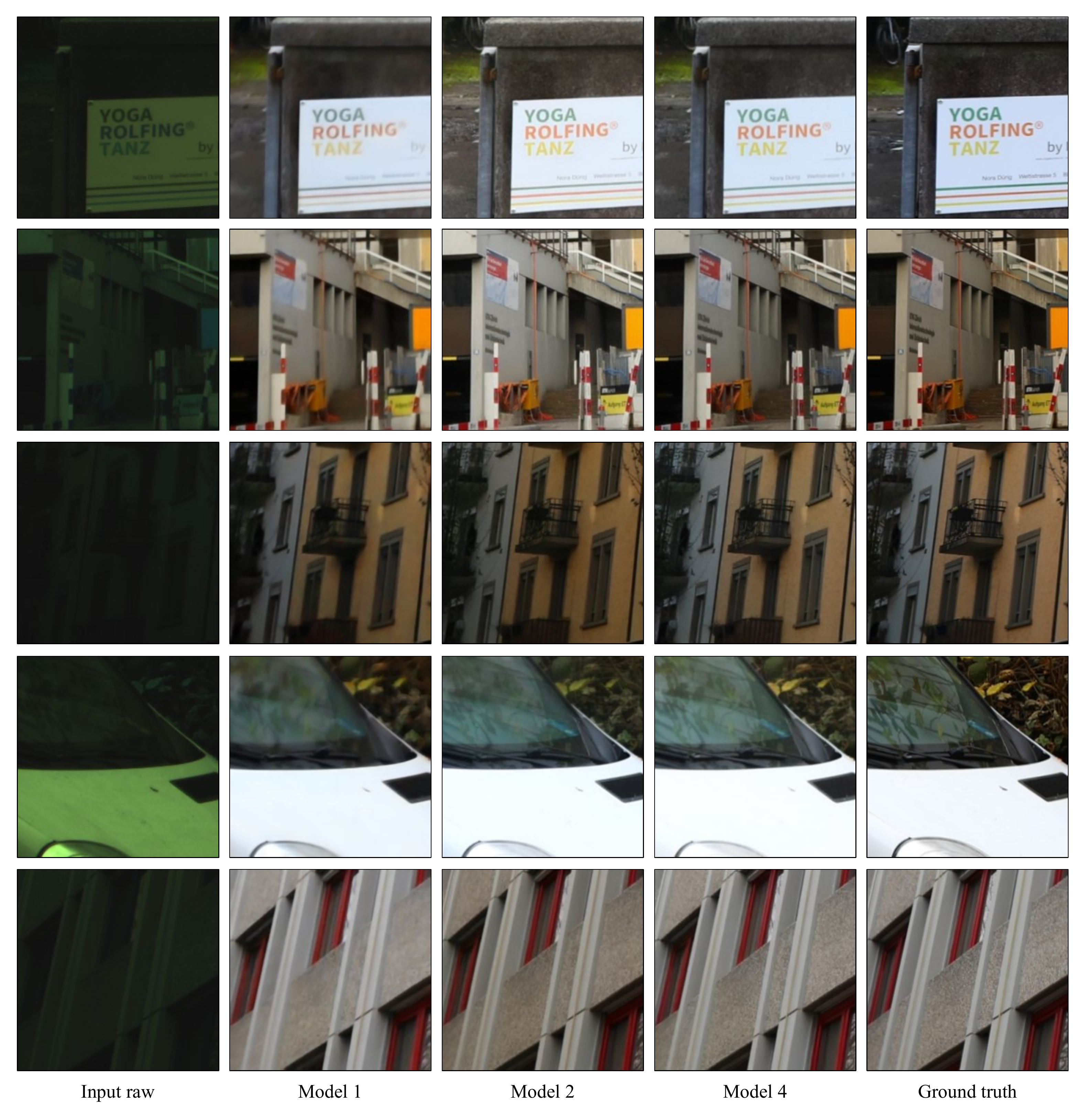}
\DeclareGraphicsExtensions.
\caption{Ablation results on different loss functions. Best viewed with zoom.}
\label{fig:3}
\end{figure*}

\paragraph{Loss function.}
Secondly, we verified the efficiency of our loss function on ZRR. Table 2 and Figure \ref{fig:3} shows the experimental results. In this experiment, the two-stage network model described in the previous section was used. As expected, the model trained using only pixel-wise L1 loss (Model 1) obtained the lowest PSNR and produced blurry results as shown in Figure \ref{fig:3}. Combining the pixel loss with the feature loss (Model 2 and Model 3) led to approximately 0.2 dB gain and successfully generated more sharp images. Note that Model 2 and Model 3 used the layers of `relu4\_1' and `relu5\_1’ of VGG-19 Network, respectively, to calculate the perceptual loss. By further incorporating the color loss (Model 4 and Model 5), 0.05dB PSNR increase was achieved and better color transformed output images were obtained as shown in Figure \ref{fig:3}. The layers of `relu4\_1' and `relu5\_1’ were used for Model 4 and Model 5, respectively. To boost the performance in the challenge, we adopted a model ensemble. As shown in Table 3, we observed that averaging the output images of Model 2, and Model 4, and Model 5 largely improves the PSNR around 0.5 dB. 

\begin{table}[t]
\renewcommand{\arraystretch}{1.2}
\label{table:2}
\begin{center}
\begin{tabular}{c|c|c|c}
\hline \hline
Model & Loss function & PSNR & SSIM\\
\hline
Model 1 & $L_{pixel}$ & 22.82 & 0.8741 \\
Model 2 & $L_{pixel}+L_{feat}$ & 23.12 & 0.8755 \\
Model 3 & $L_{pixel}+L_{feat}$ & 23.14 & 0.8709 \\
Model 4 & $L_{pixel}+L_{feat}+L_{color}$ & 23.18 & 0.8750 \\
Model 5 & $L_{pixel}+L_{feat}+L_{color}$ & 23.19 & 0.8719 \\
\hline
Ensemble &   -  & \textbf{23.70} & \textbf{0.8826} \\
\hline \hline
\end{tabular}
\end{center}
\caption{Ablation studies on loss functions}
\end{table}

\begin{table}[t]
\label{table:3}
\begin{center}
\begin{tabular}{c|c|c|c|c|c}
\hline \hline
 & \multicolumn{3}{c|}{Track 1} & \multicolumn{2}{c}{Track 2} \\
\hline
Rank & Method & PSNR & SSIM & Method & MOS\\
\hline
\textbf{1} & \textbf{W-Net} & \textbf{22.59} & \textbf{0.81} & $1^{st}$ & 1.24\\
\textbf{2} & $2^{nd}$ & 22.24 & 0.80 & \textbf{W-Net} & \textbf{1.28}\\
3 & $3^{rd}$ & 21.94 & 0.79 & $3^{rd}$ & 1.46\\
4 & $4^{th}$ & 21.91 & 0.79 & $4^{th}$ & 1.56\\
5 & $5^{th}$ & 20.85 & 0.77 & $5^{th}$ & 1.92\\
6 & $6^{th}$ & 19.46 & 0.53 & $6^{th}$ & 2.16\\
\hline \hline
\end{tabular}
\end{center}
\caption{The result of the AIM raw-to-RGB mapping challenge for the two tracks.}
\end{table}

\subsection{AIM 2019 raw-to-RGB Mapping Challenge}
\begin{figure*}[t]
\centering 
\includegraphics[width=1.0\linewidth]{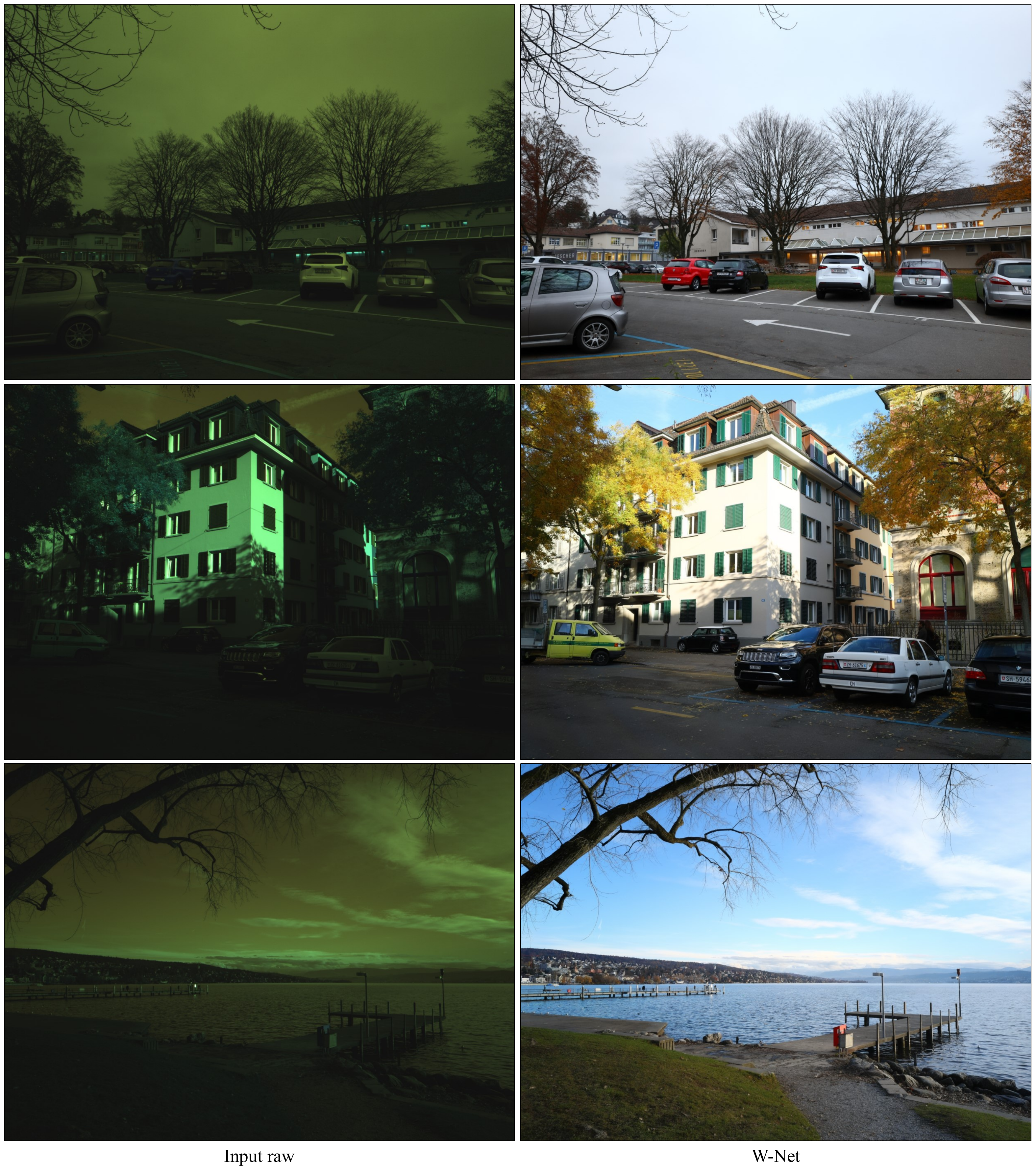}
\DeclareGraphicsExtensions.
\caption{Qualitative results of our W-Net on the Track 2 in the AIM 2019 raw-to-RGB mapping challenge. Best viewed with zoom.}
\label{fig:4}
\end{figure*}

AIM 2019 raw-to-RGB mapping challenge \cite{AIM2019RAW} consists of two tracks: fidelity track (Track 1) and perceptual track (Track 2). In the Track 1, the average PSNR and SSIM are calculated. In the Track 2, the Mean-Opinion-Score (MOS) is obtained from human subjects. Note that the full-resolution input raw images were provided for the Track 2. We submitted our ensemble model described in the previous section for the Track 1 and Model 2 for the Track 2. As shown in Table 4, our model ranked 1st place in the Track 1 and outperformed the second place by a large margin (0.35 dB). Our results for the Track 2 ranked second place, where the MOS difference between our method and the first-place method is 0.04. Qualitative results are shown in Figure \ref{fig:4}. It is observed that W-Net produces well color-transformed images with clear details.

\section{Conclusion}
We described our solution for AIM 2019 raw-to-RGB mapping challenge. To solve the challenging issues in the released dataset, we developed an effective network architecture and a loss function. We enhanced U-Net and built the two-stage network model for the task. Through the ablation studies, we verified that our loss function can handle the data misalignment and color transformation. Also, we boosted the performance by combining the models trained with different loss functions. As a result, we achieved the best quantitative results and the second-best qualitative results in the challenge.

\section{Acknowledgement}
This work was supported by Institute of Information \& Communications Technology Planning \& Evaluation (IITP) grant funded by the Korea government (MSIT) (2014-3-00077-006, Development of global multi-target tracking and event prediction techniques based on real-time large-scale video analysis).

{\small
\bibliographystyle{ieee_fullname}
\bibliography{egbib}
}

\end{document}